\documentclass[aps,prb,amsmath,amssymb,11pt]{revtex4-1}
\usepackage{epsfig,color,graphicx}
\usepackage{xr} % referencing across multiple files

\newcommand{\ket}[1]{\ensuremath{\vert #1 \rangle}}
\newcommand{\bra}[1]{\ensuremath{\langle #1 \vert}}
\newcommand{\braket}[2]{\ensuremath{\langle #1 \vert #2 \rangle}} % bra-ket inner product
\newcommand{\op}[1]{\ensuremath{\hat{#1}}} % operator
\newcommand{\trace}{{\rm Tr}}
\newcommand{\bea}{\begin{eqnarray}}
\newcommand{\eea}{\end{eqnarray}}
\newcommand{\br}{\ensuremath{\mathbf{r}}}

\newcommand{\sinfo}{Supplementary Material}
\begin{document}
\bibliographystyle{naturemag}

\title{Robust linear-scaling optimization of compact localized orbitals in density functional theory}

\author{Yifei Shi}
\author{Jessica Karaguesian}
\author{Rustam Z. Khaliullin}
\email{rustam.khaliullin@mcgill.ca}
\affiliation{Department of Chemistry, McGill University, 801 Sherbrooke St. West, Montreal, QC H3A 0B8, Canada}

%\date{\today}

\begin{abstract} 
Locality of compact one-electron orbitals expanded strictly in terms of local subsets of basis functions can be exploited in density functional theory (DFT) to achieve linear growth of computation time with systems size, crucial in large-scale simulations. However, despite advantages of compact orbitals the development of practical orbital-based linear-scaling DFT methods has long been hindered because a compact representation of the electronic ground state is difficult to find in a variational optimization procedure. In this work, we show that the slow and unstable optimization of compact orbitals originates from the nearly-invariant mixing of compact orbitals that are mostly but not completely localized within the same subsets of basis functions. 
We also construct an approximate Hessian that can be used to identify the problematic nearly-invariant modes and obviate the variational optimization along them without introducing significant errors into the computed energies. This enables us to create a linear-scaling DFT method with a low computational overhead that is demonstrated to be efficient and accurate in fixed-nuclei calculations and molecular dynamics simulations of semiconductors and insulators.
\end{abstract}
\maketitle

\section{Introduction}

At present, Kohn-Sham (KS) density functional theory (DFT) is the most popular electronic structure method. 
Unfortunately, the computational cost of the conventional diagonalization-based KS DFT grows cubically with the number of atoms preventing its application to large systems. 
To overcome this limitation, substantial efforts have been directed to the development of linearly-scaling (LS) DFT methods~\cite{goedecker1999linear,bowler2012methods,zalesny2011linear}. 

In LS DFT methods, the delocalized eigenstates of the effective KS Hamiltonian must be replaced with an alternative set of \emph{local} electron descriptors. 
Most LS methods exploit the natural locality of the one-electron density matrix (DM)~\cite{li1993density, lee1996linear, li2003density, shao2003curvy, vandevondele2012linear, kussmann2013linear, aarons2016perspective}.  
They include the Fermi operator expansion~\cite{goedecker1994efficient, goedecker1995tight}, divide-and-conquer~\cite{yang1991direct, yang1991local}, and direct DM optimization methods~\cite{li1993density, shao2003curvy, vandevondele2012linear}. 
However, the variational optimization of the DM is very inefficient for accurate DFT calculations which require many basis functions per atom~\cite{goedecker1999linear,vandevondele2012linear, arita2014stable, bowler2012methods, khaliullin2013efficient}.
Therefore, the application of DM-based LS methods have been mostly restricted to minimal-basis tight-binding problems~\cite{richters2014self, goringe1997tight, ratcliff2018band}. 
% richters2014self :   doi.org/10.1063/1.4869865
This issue is rectified in optimal-basis DM methods~\cite{gillan1998first, skylaris2005introducing, nakata2015optimized, mohr2015accurate} that contract large basis sets into a small number of new localized functions and then optimize the DM in the contracted basis. 
Despite becoming the most popular LS DFT, the efficiency of optimal-basis methods is hampered by the costly optimization of both the contracted orbitals and the DM~\cite{mostofi2003preconditioned}.

One-electron orbitals expanded strictly in subsets of localized basis functions centered on and near a given nuclei 
%(e.g. numerical atomic orbitals, Gaussian orbitals, periodic sinc or B-spline functions) 
represent another type of local electron descriptors, alternative to the DM. 
Since their introduction~\cite{matsuoka1977expansion, stoll1977on, mehler1977self} such localized orbitals have become known under different names including absolutely localized molecular orbitals~\cite{stoll1980use}, localized wave functions~\cite{ordejon1995linear}, non-orthogonal generalized Wannier functions~\cite{skylaris2005introducing}, multi-site support functions~\cite{nakata2015optimized}, and non-orthogonal localized molecular orbitals~\cite{burger2008linear}. 
In this work, we will refer to them as compact localized molecular orbitals (CLMOs) to emphasize that their expansion coefficients are zero for the basis functions outside orbital's localization subset. Unlike our previous work~\cite{khaliullin2013efficient}, we avoid using the ALMO acronym~\cite{stoll1980use} because it is commonly used now to refer to compact orbitals restricted to \emph{nonoverlapping} subsets~\cite{khaliullin2006efficient, khaliullin2007unravelling, khaliullin2008analysis, horn2013unrestricted} -- a special case of CLMOs. 

\begin{figure}
\includegraphics[width=0.5\textwidth]{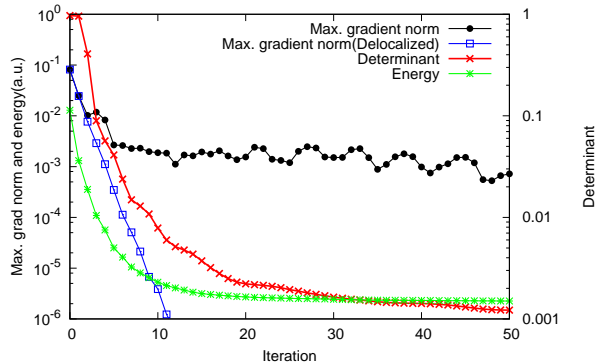}
\caption{
The determinant of the CLMO overlap matrix, the energy, and the maximum norm of the energy gradient with respect to the CLMO coefficients in a direct iterative optimization with the conjugate gradient algorithm. 
Linear system of four hydrogen fluoride molecules interacting through hydrogen bonds is described at the PBE/DZVP level of theory. 
CLMOs of a molecule are allowed to delocalize only over the two nearest neighbors. 
The zero energy is set at the energy of the fully delocalized state. 
For the CLMO state, the energy plateaus even though the gradient does not vanish. 
For the reference, the maximun norm of the gradient of same optimization using fully delocalized orbitals is also shown.}
\label{fig:det}
\end{figure}

From the computational point of view, a direct variation of CLMOs is preferable to the DM for systems with nonvanishing band gap because LS can be achieved with significantly fewer variables. 
The computational advantages of orbitals-only LS DFT would be especially pronounced in accurate calculations that require many basis functions per atom. 
CLMOs are also advantageous from the physical point of view because they provide clear, chemically meaningful, transferable description of interactions between atoms and molecules~\cite{stoll1980use, mo2000energy, khaliullin2007unravelling, khaliullin2013microscopic, mo2014block}. 
Regrettably, the development of promising orbital-based LS methods has been hindered because of slow and inherently difficult variational optimization of CLMOs~\cite{mauri1993orbital, ordejon1995linear, goedecker1999linear, fattebert2004linear, peng2013effective, tsuchida2008ab, peng2013effective}. 
In addition to poor convergence of the variational procedure, the straightforward optimization of CLMOs, which cannot be constrained to stay both compact and orthogonal~\cite{stoll1980use}, often leads to a ``collapsed'' electronic state represented by linearly dependent occupied orbitals~\cite{ordejon1995linear}. 
An example of a problematic variational optimization of CLMOs shown in Figure~\ref{fig:det} suggests that the collapse, apparent from the vanishing determinant of the CLMO overlap matrix, is closely associated with the convergences problem.

Substantial efforts have been made to mitigate these problems. 
They include the introduction of an extra set of variational parameters~\cite{yang1997absolute,burger2008linear, peng2013effective} or the energy functionals that can be calculated without inverting the CLMO overlap matrix~\cite{mauri1993orbital,kim1995total,ordejon1995linear}. 
However, despite notable progress~\cite{fattebert2004linear, fattebert2006linear, peng2013effective, osei2014accurate} the existing CLMO-based DFT algorithms 
require many more steps to reach convergence than the algorithms for fully delocalized orbitals (Figure~\ref{fig:det}).

Importantly, the CLMO convergence problem has been reliably solved for weakly-interacting molecular systems. 
It has been shown that CLMOs can be optimized efficiently only if they are forced to be orthogonal to a set of auxiliary tightly localized nonoverlaping orbitals, precomputed and fixed on their molecules~\cite{tsuchida2007augmented, tsuchida2008ab, khaliullin2013efficient}. 
As will be discussed below, this approach works only when the auxiliary orbitals resemble the final variationally optimal CLMOs and thus cannot be applied to finite-gap systems with strong covalent bonds between atoms. 

In this work, a detailed analysis of the origins of the convergence problem enabled us to develop a fast linear-scaling CLMO-based DFT method that obviates the convergence problem for systems of strongly interacting atoms and also avoids collapsed states. The proposed method is conceptually simple and does not require any precomputed tightly-localized orbitals nor the optimization of auxiliary variables. These features greatly reduce its computational cost and make it straightforward to implement. Several tests in this work demonstrate the accuracy and efficiency of the method for systems with finite band gap. This method, however, is not expected to be practical for metals and semimetals.

\section{Theory} 

\subsection{Formalism, notation and main assumptions}
 
In the first step, all atomic nuclei of a system, its electrons, and atom-centered basis set orbitals (AOs) --- in our case Gaussian functions --- are logically divided into nonoverlapping subsets called \emph{localization centers}, often referred to as \emph{fragments}. 
For a system with clearly defined molecules, a typical localization center includes all nuclei of a single molecule, associated AOs and electrons. 
For systems that cannot be partitioned into molecules --- the subject of this work --- a localization center can be represented by a single nuclei. 
As a results of the partitioning, each electron acquires a localization-center label. 
Within Kohn-Sham DFT, electrons are described by molecular orbitals (MOs) \ket{\psi_{xi}} whose indices now indicate that orbital $i$ belongs to center $x$. Throughout this work, centers are labeled with Latin letters $x,y,z$ whereas Latin letters $i,j,k$ label MOs. 

In the next step, each atom $A$ is assigned a predefined element-specific radius $R_c(A)$ that defines neighbors of each center in an obvious way: two centers are considered neighbors if there are atoms located within a sum of their radii. 
\emph{Localization domain} for each center is defined as a subset of AOs that are localized on the neighbors of a center \ket{\chi_{\bar{x}\mu}}. In our notation, index $\bar{x}$ refers to center $x$ and its neighbors. Basis set orbitals are denoted with Greek letters $\mu,\nu,\lambda, \kappa$.  By construction a basis set function may belong to several localization domains or, in other words, localization domains often overlap. 

The basis set orbitals of a localization domain form a subspace in the one-electron vector space and the following projection operator serves as the identity operator on the subspace:
\bea
\op{I}_{\bar{x}} = \ket{\chi_{\bar{x}\mu}} S^{\bar{x}\mu,\bar{x}\nu} \bra{ \chi_{\bar{x}\nu}},
\eea
where $S^{\bar{x}\mu,\bar{x}\nu}$ is a matrix element of the inverse overlap matrix $S_{\bar{x}\mu,\bar{x}\nu} = \braket{ \chi_{\bar{x}\mu}}{ \chi_{\bar{x}\nu}} $. Note that the conventional tensor notation is used to work with the nonorthogonal orbitals~\cite{head1998tensor}: covariant quantities are denoted with subscripts, contravariant quantities with superscripts, and summation is implied over the same orbital indices but not over the same center and domain indices. 

In the final step, the main approximation of CLMO methods is introduced. It is assumed that the electronic structure of the system can be described accurately by molecular orbitals that are completely localized within domains of their centers
\bea
\ket{\psi_{xi}} = \op{I}_{\bar{x}} \ket{\psi_{xi}}
\label{eq:LMO}
\eea
Thus this approximation imposes a blocked structure on the matrix of MO coefficients
\bea
\ket{\psi_{xi}} = \ket{\chi_{\bar{x}\mu}} {T^{\bar{x}\mu}}_{xi}.
\label{eq:LMOproj}
\eea
The size of blocks is determined entirely by the pre-selected $R_{c}$ and does not change with the number of atoms in the entire system. Therefore, in the limit of large systems, the computational cost of the optimization grows linearly with the number of atoms offering a way of performing LS calculation directly with MOs.

In this work, we consider only spin-unpolarized orbitals evaluated at the $\Gamma$-point. The KS DFT energy functional can be written in the conventional way 
\bea
E &=& 2 \trace \left[ \op{R} \op{H} \right] - \frac{1}{2} \int\int \frac{\rho(\br)\rho(\br')}{|\br-\br'|}d\br d\br' \nonumber \\
 &+& E_{\text{XC}} - \int v_{\text{XC}}(\br) \rho(\br) d\br,
\eea
where \op{H} is the Kohn-Sham Hamiltonian, \op{R} is the projection operator onto the occupied subspace, and $\rho(\br) = 2\bra{\br}\op{R}\ket{\br}$ is the electron density. The idempotent \op{R} is written to take into account the nonorthogonality of CLMOs
\bea \label{eq:dm}
\op{R} = \sum_{x,y} \ket{\psi_{xi}} \sigma^{xi,yj} \bra{\psi_{yj}},
\eea
where $\sigma^{xi,yj}$ is a matrix element of the inverse of the CLMO overlap matrix $\sigma_{zk,xj} = \braket{ \psi_{zk}}{ \psi_{xj}} $.

\subsection{Convergence problem and Hessian eigenspectrum}

The direct minimization of the energy functional with respect to molecular orbital coefficients is a straightforward reliable low-cost method optimization for fully delocalized orbitals ($R_c \rightarrow \infty$)~\cite{galli1992large, vandevondele2003efficient, van2002geometric} and completely localized orbitals ($R_c = 0$)~\cite{khaliullin2013efficient}. 
% voorhis2002geometric - https://doi.org/10.1080/00268970110103642
The preconditioned conjugate gradient algorithm (PCG) is a particularly efficient energy minimizer that requires iterative evaluation of the energy gradient
\bea \label{eq:grad}
{G_{\bar{x}\mu}}^{xi} \equiv \frac{\partial E}{\partial {T^{\bar{x}\mu}}_{xi}} = 4 \bra{\chi_{\bar{x}\mu}} (\op{I}-\op{R}) \op{H} \ket{\psi^{xi}}
\eea
and only a single inversion of a preconditioner. The latter is typically chosen as an easily-invertible approximation to the exact Hessian. It has been found that, for the two special cases $R_c \rightarrow \infty$~\cite{vandevondele2003efficient} and $R_c = 0$,~\cite{khaliullin2013efficient} the preconditioner defined on domain $\bar{x}$
\bea \label{eq:prec}
P_{\bar{x}\mu,\bar{x}\nu} &=& 4 \bra{\chi_{\bar{x}\mu}} (\op{I}-\op{R}) (\op{I} + \op{H})(\op{I}-\op{R}) \ket{\chi_{\bar{x}\nu}} 
\eea
provides the same rate of convergence of the PCG algorithm as the exact Hessian but at a fraction of the inversion cost. The relation between the preconditioner in Eq.~(\ref{eq:prec}) and the exact Hessian is presented in \sinfo. 

\begin{figure*}
\centering
\includegraphics[width=0.75\textwidth]{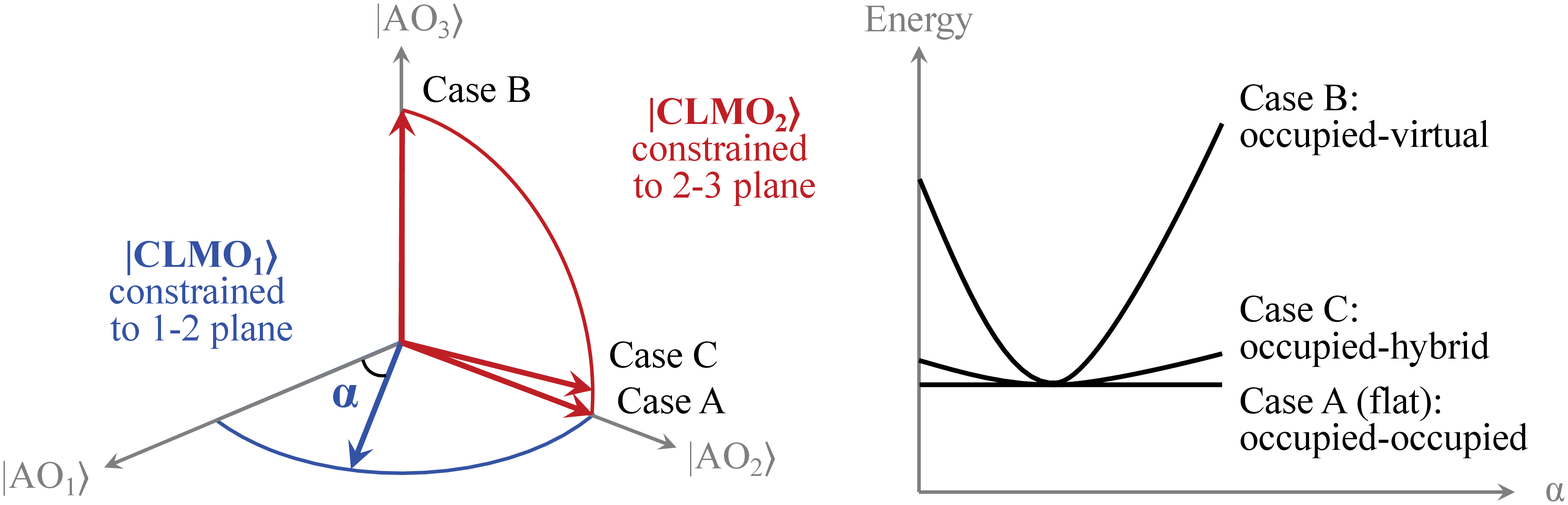}
\caption{Illustration of the origin of low-curvature modes in a model vector space spanned by three basis set functions \ket{\text{AO}_1}, \ket{\text{AO}_2}, \ket{\text{AO}_3}. The left panel shows \ket{\text{CLMO}_1} (blue) confined to its domain spanned by \ket{\text{AO}_1} and \ket{\text{AO}_2} as well as \ket{\text{CLMO}_2}  (red, three cases) confined to its domain spanned by \ket{\text{AO}_2} and \ket{\text{AO}_3}. The right panel shows the behavior of the energy as a function of the position of \ket{\text{CLMO}_1} --- angle $\alpha$ --- for three typical cases. Case C shows that the low-curvature modes arise when \ket{\text{CLMO}_2}  lies almost entirely in the domain of \ket{\text{CLMO}_1}.}
\label{fig:modes}
\end{figure*}

In the general case of finite $R_c$, the PCG-based optmization of CLMOs has been investigated thoroughly as a low-cost alternative to LS DM-based DFT methods~\cite{mauri1993orbital, kim1995total, ordejon1995linear, fattebert2004linear, fattebert2006linear, burger2008linear, peng2013effective, khaliullin2013efficient}. 
This approach is expected to be very efficient because both the CLMO coefficient and gradient matrices are small (i.e. number of columns is much smaller then the number of rows) and enforced to be sparse. 
In addition to this, the inversion of the approximate Hessian in Eq.~\ref{eq:prec} can be done fast, domain-by-domain. 
Unfortunately, the PCG algorithm, as well as all other optimization procedures (e.g. Newton-Raphson or DIIS-accelerated diagonalizations)~\cite{stoll1980use}, suffer from the aforementioned slow convergence and orbital collapse problems. 
A closer inspection of the eigenvalues of the preconditioner obtained by solving the generalized eigenproblem for each domain 
\bea
P_{\bar{x}\mu,\bar{x}\nu} {A^{\bar{x}\nu}}_{xp} =  S_{\bar{x}\mu,\bar{x}\lambda} {A^{\bar{x}\lambda}}_{xp} \Lambda_{xp},
\label{eq:gev}
\eea
reveals the origin of this and other closely related previously reported problems~\cite{goedecker1999linear}. 
The eigenvalues $\Lambda_{xp}$, which approximate the energy curvature along the optimization direction \ket{d_{xp}} represented by eigenvector ${A^{\bar{x}\nu}}_{xp} \equiv \braket{\chi^{\bar{x}\nu}}{d_{xp}}$, can be divided into three categories according to their magnitudes. 
Zero eigenvalues in the first category represent optimization directions towards occupied orbitals localized completely within the same domain (Figure~\ref{fig:modes}, case A). 
As expected the energy is invariant along these occupied-occupied mixing modes. 
The second category includes large nonzero eigenvalues that represent the optimization in the direction toward unoccupied orbitals (Figure~\ref{fig:modes}, case B) of the domain. 
These two categories are the only present in the rapidly converging optimization of fully delocalized orbitals ($R_c \rightarrow \infty$) and absolutely localized orbitals ($R_c = 0$). 
The eigenvalues classified as the third category are extremely small but nonzero (Figure~\ref{SI-sfig:hesseig}). 
They appear only for finite $R_c$ when domains share basis set functions. 
The optimization along these low-curvature nearly-invariant directions is difficult to converge because various analytical approximations (e.g. approximate Hessian, quadratic linear search) and numerical noise (e.g. finite DFT grids) make calculations  imprecise. 
In other words, optimization with finite $R_c$ becomes ill-conditioned and the low-curvature modes represents the major barrier to the practical use of promising orbital-based LS DFT methods. 

\section{Results and discussion}

\subsection{Nature of low-curvature modes} \label{marker:nature} 

What is the physical origin of the low-curvature optimization modes? 
It has been suggested previously~\cite{goedecker1999linear} that the sluggish optimization in orbital-based LS DFT is due to the inexact invariance of the energy with respect to the mixing of occupied orbitals. 
Although this might indeed present a problem for a series of functionals based on DMs that are not fully idempotent~\cite{mauri1993orbital,kim1995total,ordejon1995linear} the DM used in this work is constructed with the inverse of the overlap matrix and, therefore, is properly idempotent and exactly invariant to the mixing among occupied states. 

We suggest that the low-curvature optimization modes represent \emph{hybrid} directions, which are neither occupied nor virtual. From the point of view of the local vector subspace of $\bar{x}$, these directions exist because CLMOs of neighbor centers are only partially localized on $\bar{x}$. 
Hybrid directions, originating from CLMOs that are almost but not entirely localized on the domain, are expected to be particularly problematic (Figure~\ref{fig:modes}, case C). 

A hybrid low-curvature mode $\ket{d_{\bar{x}p}}$ is expected to have only a small component in the unoccupied subspace of domain $\bar{x}$. 
This component is measured by the residue $\Delta_{\bar{x}p}$ 
\bea
\label{eq:residue}
\Delta_{\bar{x}p} \equiv \bra{d_{\bar{x}p}} \op{I}_{\bar{x}} - \op{R}_{\bar{x}} \ket{d_{\bar{x}p}}, 
\eea
where $\op{R}_{\bar{x}}$ is a projector onto the subspace of the occupied CLMOs trimmed with operator $\op{T}_{\bar{x}} \approx \op{I}_{\bar{x}}$ to be fully localized on $\bar{x}$ 
\bea
\op{R}_{\bar{x}} &=& \sum_{y,z \in \bar{x}} \op{T}_{\bar{x}} \ket{\psi_{yi}} \sigma_{\bar{x}}^{yi,zj} \bra{\psi_{zj}} \op{T}_{\bar{x}}
%R^{\bar{x}\mu,\bar{x}\nu} &\equiv& \bra{\chi^{\bar{x}\mu}} \op{R}_{\bar{x}} \ket{\chi^{\bar{x}\nu}} \\
\label{eq:C}
\eea
and $\sigma_{\bar{x}}$ is the overlap matrix of the truncated orbitals. 

\begin{figure}
\centering
\includegraphics[width=0.5\textwidth]{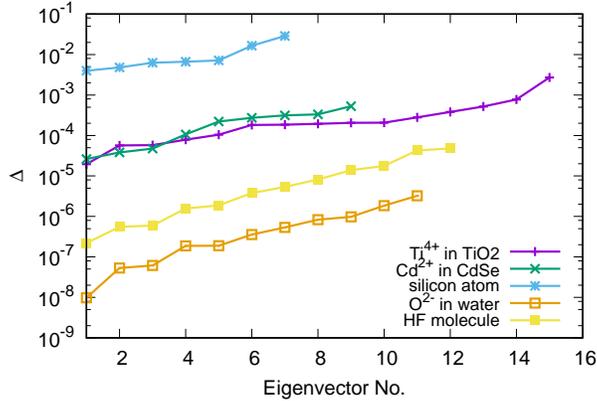}
\caption{
The norm of the projection of the small-curvature modes on the unoccupied subspace of a domain. 
Preconditioner eigenvectors with the eigenvalues smaller than 0.02~a.u. are chosen as small-curvature modes. 
The following localization ceneters are considered: Ti$^{4+}$ ions in the TiO$_2$ rutile lattice, Cd$^{2+}$ ions in the CdSe Wurtzite lattice, Si atom in the diamond silicon lattice, O$^{2-}$ in a water tetramer system, hydrogen fluoride molecule in a linear tetramer system. 
The BLYP/TZV2P level of theory is used for water and PBE/DZVP in all other tests.}
\label{fig:projection}
\end{figure}

Figure~\ref{fig:projection} shows that the small-curvature modes indeed lie mostly outside the unoccupied space for a variety of materials (Figures~\ref{SI-sfig:t_delta_tio2}--\ref{SI-sfig:t_delta_si}). 
It is also interesting to note that the unoccupied fraction of the low-curvature modes increases with the strength of interatomic delocalization (i.e. the degree of covalency of bonding). 

Thus the optimization along the low-curvature modes of a domain represents mixing of its occupied CLMO with the occupied orbitals of its neighbors that are not fully localized on the same domain. 
Unsurprisingly, such orbital variations often lead to linear dependencies among occupied orbitals and eventual collapse of the optimization. 

It should be noted that low-curvature modes also present a problem during the optimization of the support functions in optimal-basis DM methods~\cite{gillan1998first, skylaris2005introducing, nakata2015optimized, mohr2015accurate}. It is unclear whether the ``redundancy ill-conditioning'' described in Ref.~\onlinecite{gillan1998first} is of the same nature as the low-curvature modes in orbitals-only method described here.

The proposed explanation of the nature of the low-curvature modes is consistent with the fact that, for molecular partitioning, the number of the low-curvature modes is equal to the sum of occupied orbitals on the neighbor fragments. It also explains why the two-stage procedure of Ref.~\onlinecite{khaliullin2013efficient} works so well for molecular systems. In the first stage of this procedure, $R_c$ is set to zero and the resulting block-diagonal orbitals \ket{\psi_{xi}^{0}} are optimized variationally to construct the occupied space projector $\op{R}^{0}$. This projector is then fixed, $R_c$ is reset to its original finite value to allow intercenter electron delocalization, and the following trial orbitals are optimized with respect to CLMOs coefficients:
\bea
\ket{\psi_{xi}} &=& \ket{\psi_{xi}^{0}} + \op{I}_{\bar{x}} (\op{I} - \op{R}^{0} ) \ket{\chi_{\bar{x}\mu}} \bar{T}^{{\bar{x}\mu}}_{xi}
\eea
For these CLMOs, the gradient and preconditioner
%%
%\bea \label{eq:grad-bar}
%\bar{G}{_{\bar{x}\mu}}^{xi} &\equiv & \frac{\partial E}{\partial \bar{T}^{{\bar{x}\mu}}_{xi}} = 
%\bra{\chi_{\bar{x}\mu}} \op{Q}_{\bar{x}}^{0} \ket{\chi^{\bar{x}\nu}} {G_{\bar{x}\nu}}^{xi}
%\eea
%%
%
\bea \label{eq:grad-bar}
\bar{G}{_{\bar{x}\mu}}^{xi} &\equiv & \frac{\partial E}{\partial \bar{T}^{{\bar{x}\mu}}_{xi}} = 
\bra{\chi_{\bar{x}\mu}} \op{I} - \op{R}^{0} \ket{\chi^{\bar{x}\nu}} {G_{\bar{x}\nu}}^{xi}
\eea
%%
%\bea \label{eq:prec-bar}
%\bar{P}_{\bar{x}\mu,\bar{x}\nu} &=& 
%\bra{\chi_{\bar{x}\mu}} \op{Q}_{\bar{x}}^{0} \ket{\chi^{\bar{x}\lambda}}  P_{\bar{x}\lambda,\bar{x}\kappa} \bra{\chi^{\bar{x}\kappa}} \op{Q}_{\bar{x}}^{0} \ket{\chi_{\bar{x}\nu}}
%\eea
%%
%
\bea \label{eq:prec-bar}
\bar{P}_{\bar{x}\mu,\bar{x}\nu} &=& 
\bra{\chi_{\bar{x}\mu}} \op{I} - \op{R}^{0} \ket{\chi^{\bar{x}\lambda}}  P_{\bar{x}\lambda,\bar{x}\kappa} \bra{\chi^{\bar{x}\kappa}} \op{I} - \op{R}^{0} \ket{\chi_{\bar{x}\nu}} \nonumber \\
\eea
are related to those in Eqs.~(\ref{eq:grad}) and~(\ref{eq:prec}) via domain-specific operators
% $\op{I}_{\bar{x}} (\op{I} - \op{R}^{0} ) \op{I}_{\bar{x}}$.
%
\bea \label{eq:q0}
%\op{Q}_{\bar{x}}^{0} &\equiv & 
\op{I}_{\bar{x}} (\op{I} - \op{R}^{0} ) \op{I}_{\bar{x}}
\eea

As shown in Ref.~\onlinecite{khaliullin2013efficient}, these operators are essential to converging orbital optimization for molecular systems. The present work explains that this is because $\op{R}^{0}$ satisfies two important requirements. First, it is constructed from CLMOs fully localized on their centers. This ensures that each localization domain contains an integer number of occupied states. Second, $\op{R}^0$ is already close to the final converged DM for systems of weakly-interacting molecules. This, therefore, guarantees that correct low-curvature optimization modes are removed when in the second optimization stage when $\op{I}_{\bar{x}} \op{R}^{0} \op{I}_{\bar{x}}$ is projected out.  

Unfortunately, the two-stage approach does not work for systems with strong covalent bonds between localization centers (i.e. atoms). This is illustrated in Figure~\ref{fig:convergence} for the cadmium selenide lattice. The main reason for the deteriorating convergence is the inability of $\op{R}^0$, which is constructed from the block-diagonal CLMOs, to adequately represents the low-curvature modes in the delocalized states encountered in later stages of the optimization. 

\subsection{New approach to the orbital optimization}

We propose to obviate the convergence problem by detecting the low-curvature modes directly by diagonalizing the approximate Hessian in Eq.~(\ref{eq:prec}) and avoiding the optimization along these modes altogether. Although this procedure is not expected to produce fully optimized orbitals it can still find an accurate representation of the ground state. This is because the low-curvature modes are associated with mixing of occupied and nearly-occupied orbitals and, therefore, are not expected to produce a noticeable variational decrease in the energy.

To remove the low-curvature modes we construct the projector onto these modes
\bea
\op{L}_{\bar{x}} &=& \ket{d_{\bar{x}p}} \, \Theta\left[ \Lambda_{\bar{x}p} - \Lambda_c \right] \bra{d_{\bar{x}p}},
\eea
where \bra{d_{\bar{x}p}} are eigenvectors of the approximate Hessian from Eq.~(\ref{eq:gev}), $\Theta$ is the reversed unit step function 
\bea
\Theta \left[\Lambda_{\bar{x}p} - \Lambda_c \right] =
\begin{cases} 
      1 & \Lambda_{\bar{x}p} \leq \Lambda_c ,\\
      0 & \Lambda_{\bar{x}p} > \Lambda_c,
\end{cases}
\eea
and $\Lambda_c$ is the curvature threshold below which the optimization mode is classified as a low-curvature mode. The low curvature modes are then projected out from the the gradient and preconditioner in the PCG optimization algorithm
\bea \label{eq:grad-lcp} 
\tilde{G}{_{\bar{x}\mu}}^{xi} &=& \bra{\chi_{\bar{x}\mu}} \op{I} - \op{L}_{\bar{x}} \ket{\chi^{\bar{x}\nu}} {G_{\bar{x}\nu}}^{xi} 
\eea
\bea \label{eq:prec-lcp}
\tilde{P}_{\bar{x}\mu,\bar{x}\nu} &=& \bra{\chi_{\bar{x}\mu}} \op{I} - \op{L}_{\bar{x}} \ket{\chi^{\bar{x}\lambda}}  P_{\bar{x}\lambda,\bar{x}\kappa} \bra{\chi^{\bar{x}\kappa}} \op{I} - \op{L}_{\bar{x}} \ket{\chi_{\bar{x}\nu}}. \nonumber \\
\eea
As a result, the PCG optimization procedure neglects the optimization modes with the curvature below pre-selected threshold $\Lambda_c$. We will refer to the new method based on Eqs.~(\ref{eq:grad-lcp}) and (\ref{eq:prec-lcp}) as low-curvature projector (LCP) method.

\subsection{Remarks}

It is important to make several remarks about the LCP method. Although the gradient and preconditioner equations of the LCP approach resemble those of the two-stage procedure the key difference between them is that $\op{L}_{\bar{x}}$ depends on the variational parameters $T$ while $\op{I}_{\bar{x}}\op{R}^{0}\op{I}_{\bar{x}}$ does not. One of the consequences of this dependence is that there are no well-defined trial orbitals and energy functional that correspond to the LCP gradient exactly. However, the LCP gradient in Eq.~(\ref{eq:grad-lcp}) is an approximation to the gradient of the conventional energy functional with the following trial CLMOs
\bea \label{eq:telescopic-trial}
\ket{\psi_{xi}} &=& \op{I}_{\bar{x}} (\op{I} - \op{R}_{\bar{x}} ) \ket{\chi_{\bar{x}\mu}} {T^{\bar{x}\mu}}_{xi}
\eea
where $\op{R}_{\bar{x}}$ is defined in Eq.~(\ref{eq:C}) as the $T$-dependent projector onto the subspace of the CLMOs that are truncated to be fully localized on $\bar{x}$. The LCP gradient can be obtained from these trial orbitals if the $T$-dependence of $\op{R}_{\bar{x}}$ is ignored upon differentiation and $\op{R}_{\bar{x}}$ is approximated with the low-curvature projector $\op{L}_{\bar{x}}$. As described above the latter approximation makes sense because the low-curvature modes arise from the incomplete localization of the neighbors' CLMOs on $\bar{x}$. 

It is worth mentioning in passing that finding the exact gradient of the trial CLMOs in Eq.~(\ref{eq:telescopic-trial}) is tremendously difficult both analytically and algorithmically because of the cyclic dependence of $\op{R}_{\bar{x}}$ and $\ket{\psi_{xi}}$ on each other. The LCP approach described here represents a compromise between a computationally efficient and fully variational optimization.

While previous CLMO works~\cite{tsuchida2007augmented, tsuchida2008ab, khaliullin2013efficient} dealt with weakly-interacting fragments (e.g. molecules) the LCP approach represents the ultimate atomic partitioning scheme that draws fragment boundaries across strong covalent bonds. Although the LCP optimization is not fully variational it finds the orbitals and energy very close to the ground state if $\Lambda_c$ is small. 

It should be noted that any divide-and-conquer approach, including the LCP method, assigns electrons to localization centers \emph{a priori} and, therefore, requires at least a rough idea about the optimal electron distribution. Although the LCP procedure presented here is not applicable to systems with completely unknown bonding properties this issue can be resolved by employing the global optimizer described in Ref.~\onlinecite{kim1995total} in conjunction with the current approach.

\subsection{Implementation}

The procedure for finding and projecting out the low-curvature modes is implemented in the CP2K software package~\cite{kuhne2020cp2k}. CP2K relies on the mixed Gaussian and plane wave representation of the electronic degrees of freedom~\cite{vandevondele2005quickstep} and is an ideal platform for the new orbital-based LS method: just a few tightly-localized Gaussian AOs can provide an accurate representation of CLMOs, whereas plane waves ensure a fast LS construction of the KS Hamiltonian for large systems.

All matrix multiplications are performed with the DBCSR library~\cite{borvstnik2014sparse} designed for massively-parallel linear-scaling handling of large sparse matrices. 
A special care is taken to reduce the computational overhead of the optimization procedure for large Gaussian basis sets. 
To this end, the order of matrix multiplications in Eqs.~(\ref{eq:grad-lcp}) and (\ref{eq:prec-lcp}) is chosen to avoid steps that scale cubically with the size of the Gaussian basis set. 
The diagonalization of the preconditioner matrices is done independently for each domain. 

It is important to note that the construction of the DM requires the inversion of the CLMO overlap matrix. 
This matrix is small and its size is independent of the size of the basis set. 
However, it is not confined to individual domains. This inversion is carried out using the iterative Hotelling method~\cite{hotelling1943some} that is based entirely on matrix multiplications and becomes LS when the system is large and the CLMO overlap and its inverse are sparse.

\begin{figure}
\centering
\includegraphics[width=0.5\textwidth]{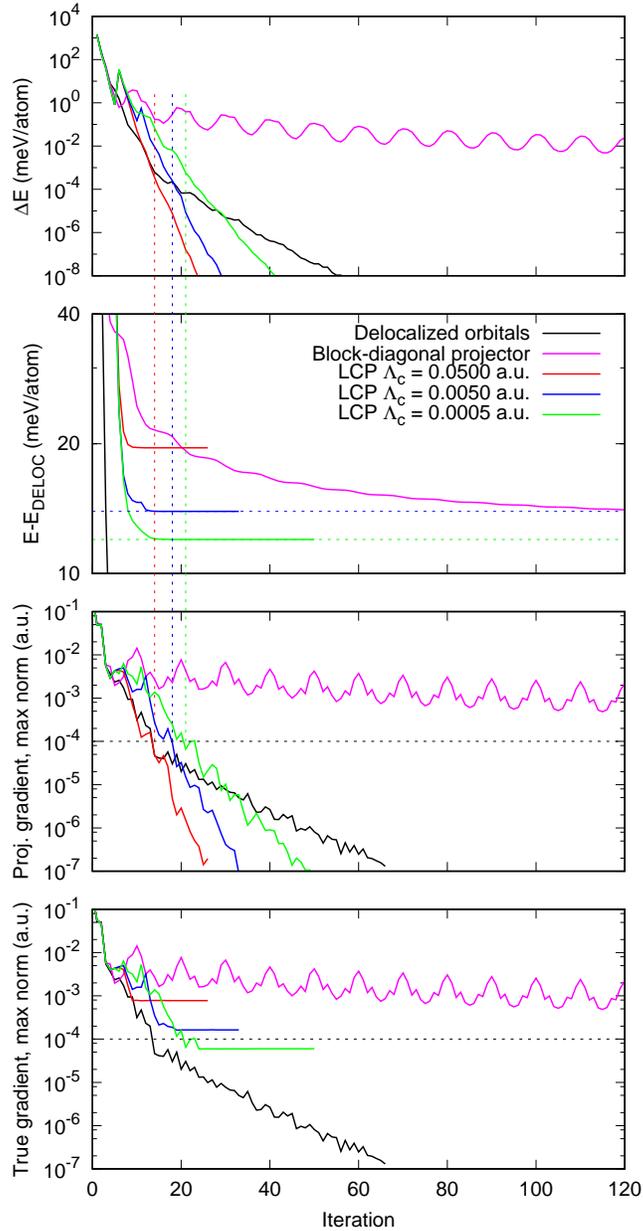}
\caption{Optimization of the CLMOs for the hexagonal wurtzite CdSe lattice. PBE/DZVP, $R_c = 4.9$~{\AA}. The block-diagonal projector is described in Eq.~\ref{eq:q0}. The $\Lambda_c = 5 \times 10^{-5}$~a.u. optimization fails.} 
\label{fig:convergence}
\end{figure}

\subsection{Accuracy and efficiency tests}

%\subsection{Convergence}

The LCP has been tested on a variety of systems, in which each fragment can be represented with closed-shell (i.e. doubly-occupied) orbitals.

Figure~\ref{fig:convergence} shows the LCP orbital optimization for the hexagonal CdSe -- a challenging case for DM-based LS methods because of the small band gap of this material. % and strong interactions between its atoms. 
Since the difficult optimization modes are neglected in the LCP optimization, the norm of the projected gradient decreases fast.  While lowering $\Lambda_c$ makes the convergence somewhat slower, the rate of the LCP gradient decrease is the same as that for the straightforward optimization of unconstrained delocalized orbitals.  
In contrast, the gradient norm decreases much slower in the two-stage block-diagonal projector optimization described for molecular systems in Ref.~\cite{khaliullin2013efficient}.
While the high rate of the LCP gradient convergence is expected (because of the explicit projection), it is crucial that the LCP energies plateau at the levels that are difficult to achieve with the block-diagonal projector (dashed lines in Figure~\ref{fig:convergence}) and impossible to reach without any regularization. For example, 14~meV/atom energy above the delocalized state is achieved in 18 iterations with the $\Lambda_c = 5 \times 10^{-3}$~a.u. LCP method and in 131 iterations with the block-diagonal projector method. The 12~meV/atom energy above the delocalized state is achieved in 21 iterations with the $\Lambda_c = 5 \times 10^{-4}$~a.u. LCP method and could not be achieved even after 500 iterations with the block-diagonal projector and trust-region methods (Figure~\ref{SI-sfig:cdse-conv}). It should be noted that $\Lambda_c < 5 \times 10^{-5}$~a.u. is too low for the LCP regularization to remain efficient and the optimization behaves similarly to the unregularized case and fails after several iterations.

The proximity of the LCP energies to the best estimates of the variational CLMO energy (i.e. underconverged block-diagonal projector optimization) indicates that the approximate Hessian, proposed to detect low-curvature modes, is sufficiently accurate and does not eliminate any important modes.
This example also shows that the LCP method does not only make the norm of the projected gradient low, it can also make the norm of the true gradient lower than that achievable with block-diagonal projector (bottom panel, Figure~\ref{fig:convergence}). While the LCP method cannot be used to make the norm of the unprojected gradient arbitrarily low (that is, it does not solve the convergence problem completely), it represents a simple regularization approach (i.e. speeds up the optimization, prevents failures) and also provides a better stopping criterion than the rate of energy change, currently used in the optimization of strictly localized orbitals (top panel, Figure~\ref{fig:convergence}). 

%\subsection{Accuracy} 

\begin{figure*}
\centering
\includegraphics[width=0.90\textwidth]{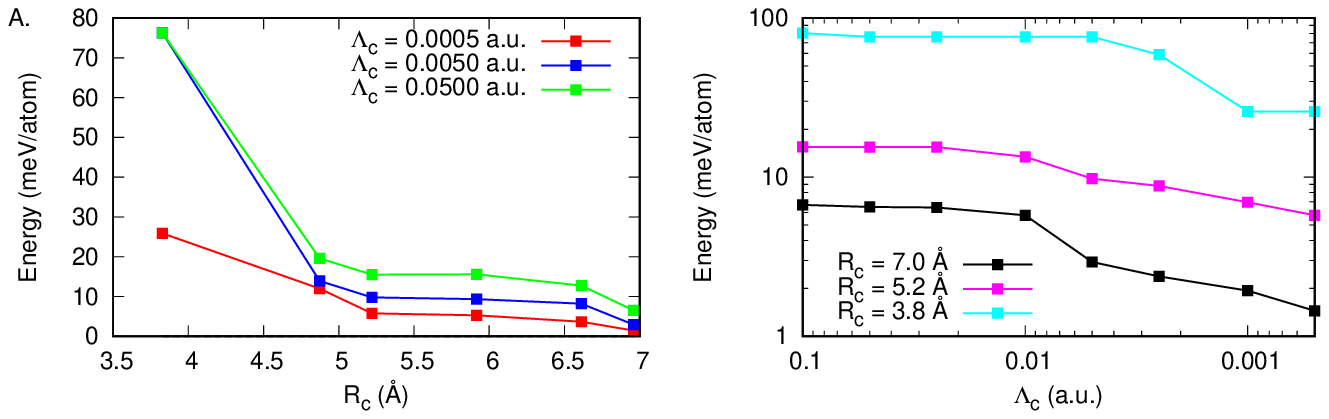}
\includegraphics[width=0.90\textwidth]{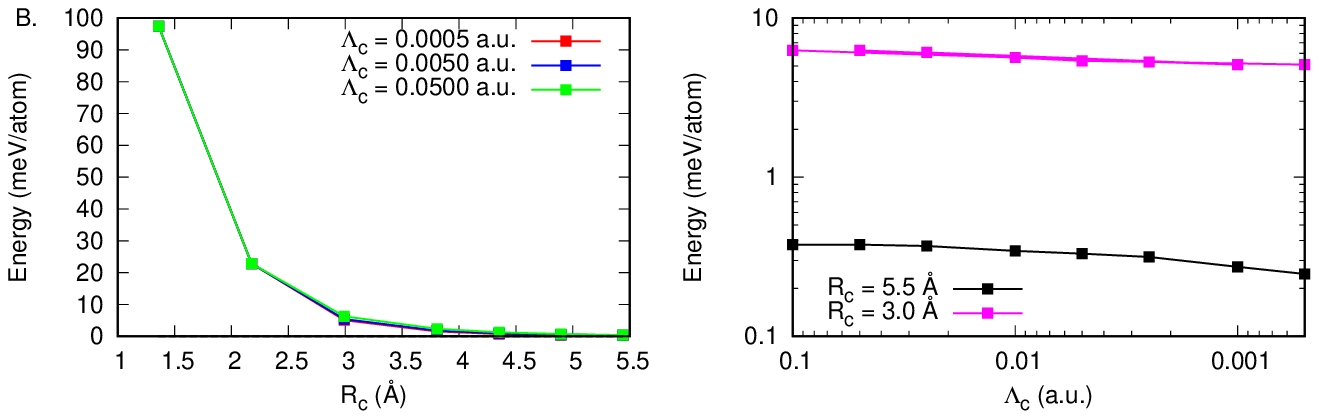}
\includegraphics[width=0.90\textwidth]{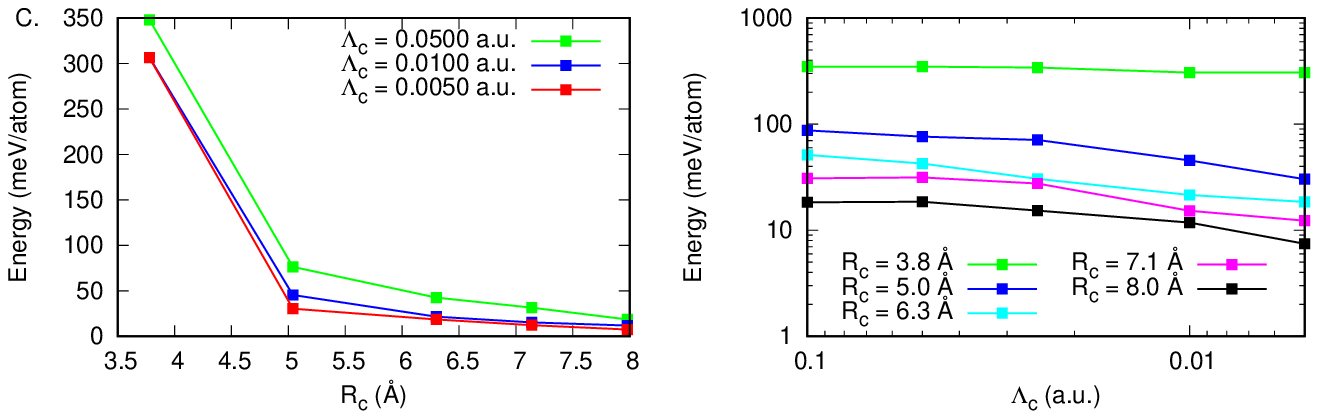}
\caption{Dependence of the LCP energy on the localization radii $R_c$ and LCP threshold $\Lambda_c$ for: A. Hexagonal wurtzite phase of CdSe (PBE/DZVP). B. Atomically-partitioned liquid water (BLYP/TZV2P). C. Cubic diamond phase of silicon (PBE/DZVP). The energy of the delocalized orbitals is chosen as zero. The convergence criteria is $\vert\vert \tilde{G} \vert\vert_{\text{max}} < 10^{-4}$~a.u.}
\label{fig:accuracy}
\end{figure*}

The accuracy of the LCP energies as a function of $R_c$ is shown in Figure~\ref{fig:accuracy} for several systems with significant covalent bonding (i.e. electron delocalization) between fragments: liquid water in which each atom is treated as a fragment, silicon in the cubic diamond lattice, and CdSe in the hexagonal wurtzite lattice.
Figure~\ref{fig:accuracy} demonstrates that the LCP energy converges to the energy of the delocalized orbtals as $R_c$ increases. It should be noted that many chemical processes in these systems can be reproduced correctly as long as $R_c$ includes the (next-)nearest neighbors and as long as the number of neighbors remain similar during a simulation of the reaction. This means that it is often not necessary to set $R_c$ to large values with the goal of reproducing the reference delocalized-state energy quantitatively.

Figure~\ref{fig:accuracy} makes it clear that $\Lambda_c$ can affect the LCP energy as much as $R_c$ and, for some systems, it is important to tune $\Lambda_c$ for best accuracy-performance compromise for every material and $R_c$. For example, systems with lower $R_c$ are influenced less strongly by $\Lambda_c$ because the number of basis set functions in the localization domains is smaller and there are fewer Hessian eigenvalues. It is worth noting that the lowest $\Lambda_c$ that allows to convergence SCF for Si ($\Lambda_c \approx 5 \times 10^{-3}$~a.u.) is different from that for CdSe ($\Lambda_c \approx 5 \times 10^{-4}$~a.u.).

\begin{figure}
\centering
\includegraphics[width=0.45\textwidth]{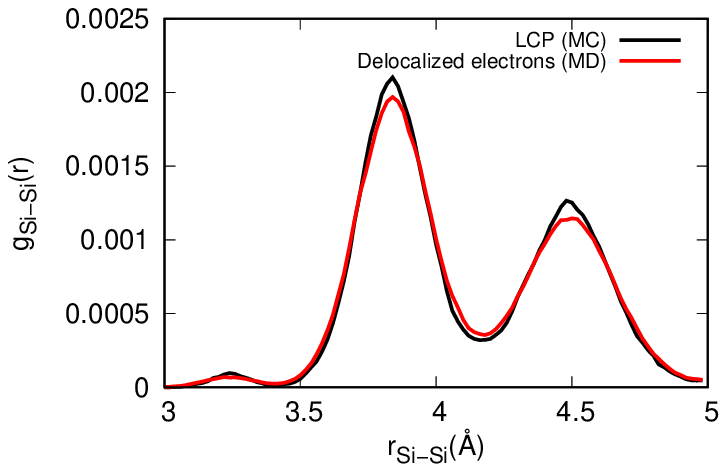}
\caption{Silicon-silicon RDF calculated for the cubic diamond phase of silicon with carbon atoms introduced to create a point defect. XC/basis simulations are performed at $T=500$~K with the conventional orbital-transformation~\cite{vandevondele2003efficient} method for the delocalized electrons and the LCP method with $R_c = 6$~\AA\ and $\Lambda_c = 0.02$~a.u. The system includes 62 silicon atoms and two carbon atoms.}
\label{fig:mc}
\end{figure}

To demonstrate the accuracy of the LCP energies further, we performed a 500~K Monte Carlo (MC) simulation of silicon in the cubic diamond lattice with a defect created by replacing two neighboring silicon atoms with two carbon atoms. Figure~\ref{fig:mc} shows that the silicon-silicon radial distribution function (RDF) at equilibrium almost perfectly reproduces that calculated using conventional DFT methods for fully delocalized electrons. 

%\subsection{Stable molecular dynamics simulations}

\begin{figure}
\includegraphics[width=0.48\textwidth]{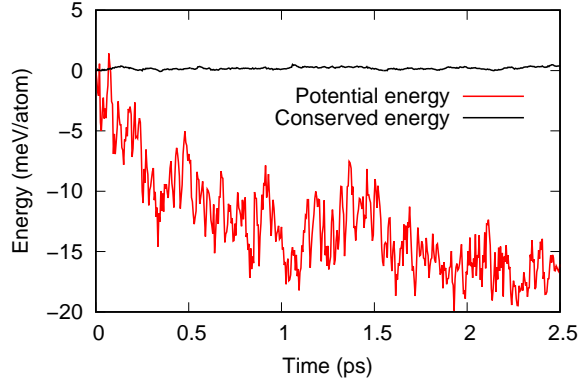}
\caption{The potential energy and the conserved quantity in a LCP-based MD simulation of a protonated water nanocluster $\text{(H}^{+}\text{)}_2\text{(H}_2\text{O)}_{62}$. The temperature of the NVT simulations is set to 298~K and controlled by a canonical velocity re-scaling thermostat~\cite{bussi2007canonical} with the coupling time constant of 50~fs. The potential and conserved energies are shifted for clarity.}
\label{fig:md}
\end{figure}

Molecular dynamics (MD) simulations present an even more stringent test on the accuracy of the LCP method. While minor energy errors are not crucial in fixed-nuclei calculations, geometry optimizations, and Monte-Carlo sampling they tend to accumulate in MD trajectories leading to non-physical sampling and eventual failure of simulations. Accurate calculation of forces is therefore crucial for molecular dynamics simulations, the stability of which are judged by the accumulated drift in the conserved quantity (e.g. total energy of the system). Although CLMOs obtained with the LCP method are not strictly variational we choose to neglect the errors associated with underconverged orbitals and invoke the Hellmann--Feynman theorem~\cite{feynman1939forces} in the calculation of atomic forces. This procedure is used to perform an LCP-based MD simulation of a protonated water nanocluster containing 62 water molecules and two protons. In the course of a 2.5~ps NVT MD simulation, the two protons hop around breaking and forming covalent bonds with water molecules. An unequilibrated simulation is shown intentionally because it contains frequent proton hopping events, that are challenging to describe correctly with atomically-partitioned CMOs. To reproduce the motion of protons around the nanodroplet each atom had to be treated as a localization center. Setting $R_c = 4.0$~\AA\ and $\Lambda_c = 0.005$~a.u. produces sufficiently optimized CLMOs to ensure the stability of the MD simulation on picosecond timescales. Figure~\ref{fig:md} shows that the drift and fluctuations in the conserved quantity remain constant compared to the fluctuations in the potential energy, despite multiple kinks arising from the sharp spatial orbital cutoff that lead to discontinuities in the potential energy surface. The applicability of the LCP method in MD simulations will be explored further in the future especially in conjunction with the modified Langevin integrator method~\cite{scheiber2018communication}. 

%\subsection{Computational efficiency}

\begin{figure}
\centering
\includegraphics[width=0.45\textwidth]{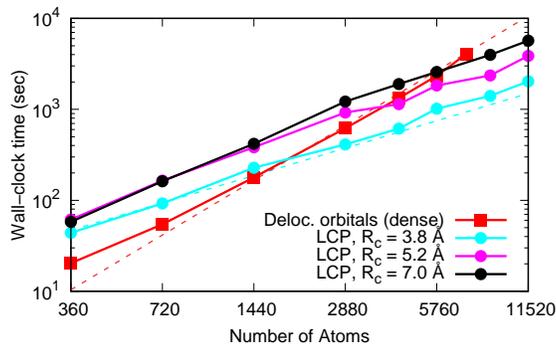}
\caption{Timing benchmark for the hexagonal phase of CdSe described at the PBE/DZVP level of theory on 400 compute cores. In LCP calculations, $\Lambda_c = 0.005$~a.u. Dashed lines show perfect linear and quadratic scaling. Note that with the employed settings, OT calculations cannot be performed due to the high memory demands of handling delocalized molecular orbitals.}
\label{fig:scaling}
\end{figure}

To test the computational efficiency of the newly designed method, we compared its performance to the orbital transformation (OT) method~\cite{weber2008direct,vandevondele2003efficient} --- a well-optimized low-prefactor cubic scaling DFT method for conventional fully delocalized orbitals. The benchmark calculations are performed for the hexagonal CdSe system of different sizes. Figure~\ref{fig:scaling} demonstrates that the implementation of the LCP method is asymptotically linear scaling. The LS regime is achieved when the MO overlap matrix and its inverse are sparse. 
The LCP calculations ($\Lambda_c = 0.005$~a.u.) become faster than the calculations with the delocalized orbtials in the region of 1500--6000 atoms. Figure~\ref{SI-sfig:scaling} shows that it is mostly the handling of sparse matrices in the LCP calculations (DBCSR library~\cite{borvstnik2014sparse}) that is responsible for the higher computational overhead compared to the calculation with delocalized orbitals (ScaLAPACK library). The precise position of the crossover point depends noticeably on the chosen localization radius: 1500, 4000 and 6000 atoms for $R_c=3.8, 5.2, 7.0$~\AA, respectively. Static calculations (e.g. single-energy, geometry optimization) with that many atoms can be considered routine today and can immediately benefit from employing the LCP method. In contrast, DFT-based simulations of thousands of atoms that require extensive sampling (e.g. MC, MD) are still considered state-of-the-art and will only benefit from the LCP technique when faster computer platforms become routinely available in the future.

\section{Conclusions} 

This work presents a new linear-scaling algorithm for the optimization of strictly localized molecular orbitals in DFT. 
In this approach, the optimization of the density matrix is not performed, resulting in low computational cost. 
The convergence problem that has plagued orbital-based LS DFT methods is eliminated by finding and obviating the optimization of orbitals along low-curvature modes -- the directions associated with tiny eigenvalues of the electronic Hessian. It is shown that an approximate Hessian can be used to detect the low-curvature modes instead of the exact electronic Hessian, the calculation and diagonalization of which is unfeasible.
It is also shown that the low-curvature modes -- an unavoidable consequence of imposing localization constraints -- can be safely neglected because they are associated with the mixing of nearly occupied local states and do not produce a noticeable variational decrease in the energy. The main practical shortcoming of the method is that the low-curvature threshold $\Lambda_c$ has to be tuned beforehand for each system to achieve best compromise between performance and accuracy. Poor choices of $\Lambda_c$ result in sluggish convergence ($\Lambda_c$ is too low) or in inaccurate energies ($\Lambda_c$ is too high).

The new methodology, which is expected to be applicable to systems with nonvanishing band gap, is tested on a variety of materials including atomically-partitioned liquid water, hexagonal phase of cadmium selenide, and cubic diamond phase of silicon with and without defects. 
These tests demonstrate that the method is accurate and efficient even when localization centers are represented by single atoms and there are strong covalent interaction between the centers. 
Furthermore, preliminary tests on protonated water nanoclusters suggest that the atomic partitioning does not present problems for the new method and it is sufficiently robust to enable stable molecular dynamics simulation of bond-breaking and formation processes. 

The developed LS DFT method is expected to have a significant impact on computational modeling of large complex systems. Due to its low computational overhead, the method will enable DFT calculations on previously inaccessible length scales making completely new chemical phenomena amenable to simulations.

\section{\sinfo}

The \sinfo\ contains the description of the relation between the exact electronic Hessian and preconditioner, eigenvalues of the preconditioner for the hexagonal phase of CdSe, illustrations of origins of the low-curvature modes, performance of the dogleg trust region algorithm, and extended timing benchmark for the hexagonal phase of CdSe.

\section{Acknowledgments} 

The research was funded by the Natural Sciences and Engineering Research Council of Canada through the Discovery Grant (RGPIN-2016-05059). The authors are grateful to Compute Canada for computer time.

\section{Data availability}

The data that support the findings of this study are available from the corresponding author upon reasonable request.

\end{document}